%% file: main_submit.tex
\documentclass[a4paper,10pt,twoside]{cpc-hepnp}

\usepackage{CJK,upgreek,fancyhdr}
\usepackage{subfigure}
\usepackage{multicol}
\usepackage{array}
\usepackage{units}
\usepackage{booktabs}
\usepackage{amsmath}
\usepackage{amsfonts,amssymb,bm,mathrsfs,bbm,amscd}
\usepackage{lastpage}
\usepackage{multirow}
\usepackage{stfloats}
\usepackage{graphicx}
\DeclareGraphicsExtensions{.pdf,.jpeg,.png}
\usepackage{epstopdf}
\usepackage[english]{babel}
\usepackage[colorlinks,urlcolor=blue,linkcolor=blue,anchorcolor=blue,citecolor=blue]{hyperref}
\usepackage{overpic}
\usepackage[title]{appendix}
\usepackage{lineno}
\usepackage{color}
\usepackage{diagbox}
\usepackage{epsfig}
\usepackage{bigstrut}
\usepackage{dcolumn}
\usepackage{xspace}
\usepackage{verbatim}
\usepackage{footmisc}
\usepackage{xcolor}
\usepackage{soul}
\usepackage{caption}
\usepackage{float}

\newcommand{\ifb}{\mathrm{fb}^{-1}}
\newcommand{\mev}{\,\mathrm{MeV}}

\newcommand{\gev}{\,\mathrm{GeV}}

\let\oldendthebibliography\endthebibliography \def\endthebibliography{\unskip.\oldendthebibliography}

\newcolumntype{d}[1]{D{.}{.}{#1}}

\begin{document}
\begin{CJK*}{UTF8}{gkai}
%%%%%%%%%%%%%%%%%%%%%%%%%%%%%%%%%%%%%%%%%%%%%%%%%%%%%%%%%%%%%
%% note here we use UTF8 and gbsn(simpe chinese character)
%% however, in author_list_en_cn.tex, there is a traditional chinese character for the name
%% S.~L.~Olsen(??)
%% so in there, we use the {\CJKfamily{bsmi}?? ??} to show that name
%% and for the name ???theÔ?Õcan not be shown in gbsn,
%% and for the name ???? {\CJKfamily{bsmi}?}
%% so we also use {\CJKfamily{bsmi}?} to show it
%% If editors want to change the font, please notice that problem.
%%%%%%%%%%%%%%%%%%%%%%%%%%%%%%%%%%%%%%%%%%%%%%%%%%%%%%%%%%%%

\fancyhead[c]{\small Chinese Physics C~~~Vol. 49, No. 8 (2025) 083001}
\fancyfoot[C]{\small 083001-\thepage}
\footnotetext[0]{Received 20 March 2025; Accepted 17 April 2025; Published online 18 April 2025}

\title{Search for radiative leptonic decay \texorpdfstring{$D^+\to\gamma e^+\nu_e$}{Dp -> gamma e nu} using deep learning\footnote{The BESIII Collaboration thanks the staff of BEPCII and the IHEP computing center for their strong support. This work is supported in part by National Key R\&D Program of China under Contracts Nos. 2020YFA0406400, 2023YFA1606000, 2020YFA0406300; National Natural Science Foundation of China (NSFC) under Contracts Nos. 11635010, 11735014, 11935015, 11935016, 11935018, 12025502, 12035009, 12035013, 12061131003, 12192260, 12192261, 12192262, 12192263, 12192264, 12192265, 12221005, 12225509, 12235017, 12361141819; the Chinese Academy of Sciences (CAS) Large-Scale Scientific Facility Program; the CAS Center for Excellence in Particle Physics (CCEPP); Joint Large-Scale Scientific Facility Funds of the NSFC and CAS under Contract No. U1832207; CAS under Contract No. YSBR-101; 100 Talents Program of CAS; CAS Project for Young Scientists in Basic Research No. YSBR-117; The Institute of Nuclear and Particle Physics (INPAC) and Shanghai Key Laboratory for Particle Physics and Cosmology; Agencia Nacional de Investigaci\'on y Desarrollo de Chile (ANID), Chile under Contract No. ANID PIA/APOYO AFB230003; German Research Foundation DFG under Contract No. FOR5327; Istituto Nazionale di Fisica Nucleare, Italy; Knut and Alice Wallenberg Foundation under Contracts Nos. 2021.0174, 2021.0299; Ministry of Development of Turkey under Contract No. DPT2006K-120470; National Research Foundation of Korea under Contract No. NRF-2022R1A2C1092335; National Science and Technology fund of Mongolia; National Science Research and Innovation Fund (NSRF) via the Program Management Unit for Human Resources \& Institutional Development, Research and Innovation of Thailand under Contract No. B50G670107; Polish National Science Centre under Contract No. 2019/35/O/ST2/02907; Swedish Research Council under Contract No. 2019.04595; The Swedish Foundation for International Cooperation in Research and Higher Education under Contract No. CH2018-7756; U. S. Department of Energy under Contract No. DE-FG02-05ER41374.}}
\maketitle
\begin{center} 
\input{authorlist_enzh_2024-11-18}
\end{center}

%\linenumbers
\begin{abstract}
Using 20.3~$\ifb$ of $e^+e^-$ annihilation data collected at a center-of-mass energy of 3.773~$\gev$ with the BESIII detector, we report on an improved search for the radiative leptonic decay $D^+\to\gamma e^+\nu_e$. An upper limit on its partial branching fraction for photon energies $E_\gamma>10~\mev$ was determined to be $1.2\times10^{-5}$ at a 90\% confidence level; this excludes most current theoretical predictions. A sophisticated deep learning approach, which includes thorough validation and is based on the Transformer architecture, was implemented to efficiently distinguish the signal from massive backgrounds.
\end{abstract}

\begin{keyword}
Charmed hadron, Radiative leptonic decay, BESIII experiment, Deep learning
\end{keyword}

\begin{multicols}{2}
%%%%%%%%%%%%%%%%%%%%%%%%%%%%%%%%%%%%%%%%%%%%%%%

\section{Introduction}
The purely leptonic decays of charmed mesons, $D^+\to l^+\nu_l\ (l=e,\ \mu,\ \tau)$, offer an ideal laboratory for testing the Standard Model. These decays also offer one of the simplest and best understood probes of the $c\to d$ quark flavour-changing transition~\cite{Silverman:1988gc,Rosner:1990xx,Chang:1992bb,Chang:1992pt}. 
However, these decays are subject to helicity suppression, with decay rates proportional to the square of the lepton mass. Consequently, the $D^+\to e^+\nu_e$ decay rate is highly suppressed compared to its muonic~\cite{HFLAV:2022esi,BESIII:2013iro} and tauonic~\cite{HFLAV:2022esi,BESIII:2019vhn} counterparts, with a branching fraction (BF) expected to be approximately $10^{-8}$, beyond the current experimental sensitivity~\cite{CLEO:2008ffk,dp2enu}. 

In the radiative leptonic decay $D^+\to\gamma e^+\nu_e$, photon emission mitigates helicity suppression, with the strong interaction engaged solely in a single hadronic external state. Theoretically, its differential decay width can be expressed as a function of the photon energy $E_\gamma$ in the rest frame of the $D^+$ meson~\cite{Lu:2021ttf} via
\begin{linenomath*}
\small
\begin{equation}
\begin{aligned}
\frac{{\rm d}\Gamma}{{\rm d}E_\gamma}(D^+\to \gamma l^+\nu_l)=&\frac{\alpha_{em}G^2_F\left[F_V^2(E_\gamma)+F_A^2(E_\gamma)\right]}{6\pi^2}\\
&\times\left|V_{cd}\right|^2E_\gamma^3m_{D^+}\left(1-\frac{2E_\gamma}{m_{D^+}}\right),
\label{eq:radlepdec}
\end{aligned}
\end{equation}
\end{linenomath*}
where $\alpha_{em}$ is the fine-structure constant, $G_F$ is the Fermi coupling constant, $\left|V_{cd}\right|$ is the Cabibbo-Kobayashi-Maskawa matrix element~\cite{Cabibbo:1963yz,Kobayashi:1973fv}, $m_{D^+}$ is the nominal mass of the $D^+$ meson~\cite{pdg}, and $F_V$ and $F_A$ are the form factors characterizing the non-perturbative quantum chromodynamics (QCD) dynamics. Various theoretical models have been proposed to investigate the $D^+\to\gamma e^+\nu_e$ decay, including quark models~\cite{Geng:2000if,Lu:2002mn,Barik:2009zza}, perturbative QCD~\cite{Korchemsky:1999qb}, lattice QCD~\cite{Desiderio:2020oej}, and QCD factorization approaches~\cite{Yang:2014rna,Yang:2016wtm,Lu:2021ttf}. Most theoretical predictions on its BF, $\mathcal{B}(D^+\to\gamma e^+\nu_e)$, are on the order of $10^{-5}$, as summarized in Table~\ref{tab:BF}.

\begin{center}
\footnotesize
\tabcaption{Theoretical predictions and experimental result for $\mathcal{B}(D^+\to\gamma e^+\nu_e)$.}
\label{tab:BF}
\begin{tabular}{c|c}
\hline\hline
Model & $\mathcal{B}\ (\times10^{-5})$\\ \hline
Light front quark model~\cite{Geng:2000if} & 0.69\\
Non-relativistic quark model~\cite{Lu:2002mn} & 0.46\\
Relativistic quark model~\cite{Barik:2009zza} & 3.34\\ 
Perturbative QCD~\cite{Korchemsky:1999qb} & $8.2\pm6.5$\\
Lattice QCD~\cite{Desiderio:2020oej} & $0.09\pm0.04$\\
QCD factorization ~\cite{Yang:2014rna} & 2.81\\
QCD factorization ~\cite{Yang:2016wtm} & 1.92\\
QCD factorization ~\cite{Lu:2021ttf} & $\left(1.88^{+0.36}_{-0.29},\, 2.31^{+0.65}_{-0.54}\right)$\\
\hline
BESIII 2017~\cite{BESIII:2017whk} & $<3.0$\\
%This work & $<1.2$ \\
\hline\hline
\end{tabular}
\end{center}

The only experimental result of $D^+\to\gamma e^+\nu_e$ was reported from the BESIII experiment~\cite{BESIII:2017whk}, based on 2.9~$\rm fb^{-1}$ of $e^+e^-$ annihilation data taken at the center-of-mass energy $\sqrt{s}=3.773~\gev$. The upper limit on the partial BF of $D^+\to\gamma e^+\nu_e$ with $E_\gamma>10$~MeV was determined to be $3.0\times10^{-5}$ at 90\% confidence level (C.L.), approaching the range of theoretical predictions~\cite{Barik:2009zza,Korchemsky:1999qb,Yang:2014rna,Yang:2016wtm,Lu:2021ttf}. 
A larger dataset comprising 20.3~$\rm fb^{-1}$ of $e^+e^-$ annihilation recently collected at BESIII~\cite{Ablikim:2013ntc,BESIII:2015equ,lumi}, which includes the previous 2.9 ~$\rm fb^{-1}$, 
provides an excellent opportunity to further constrain these theoretical models.

The major challenge in searching for $D^+\to\gamma e^+\nu_e$ is the substantial background contamination from $D^+$ semileptonic decays, such as $D^+\to\pi^0e^+\nu_e$, $D^+\to K^0_Le^+\nu_e$ and $D^+\to K^0_S(\to\pi^0\pi^0)e^+\nu_e$. Photons originating from $\pi^0$ decays or long-lived neutral hadrons can deposit energy in the electromagnetic calorimeter, potentially mimicking the radiative photon signal. Conventional cut-based methods have struggled to establish effective discriminators to suppress these backgrounds, whose yields are two orders of magnitude greater than the potential signal.

An unprecedentedly powerful signal identification tool can be accomplished using deep learning, utilizing a deep neural network (DNN), to distinguish signals from backgrounds. Leveraging flexible data representation and modern algorithms, this method exhibits powerful capabilities in recognizing and interpreting underlying relationships and hidden patterns within the topological structure of $e^+e^-$ annihilation events. 

In this paper, we present an improved search for $D^+\to\gamma e^+\nu_e$ based on the 20.3~$\ifb$ data, enhancing signal sensitivity through a deep-learning-based signal identification method. Throughout this paper, charge conjugation is implied by default.

\section{BESIII detector and Monte Carlo simulation}

The BESIII detector~\cite{BESIII:2009fln} records symmetric $e^+e^-$ collisions 
provided by the BEPCII storage ring~\cite{Yu:IPAC2016-TUYA01} in the center-of-mass energy range from 1.84 to 4.95~GeV, with a peak luminosity of $1.1 \times 10^{33}\;\text{cm}^{-2}\text{s}^{-1}$ achieved at $\sqrt{s} = 3.773\;\text{GeV}$. BESIII has collected large data samples in this energy region~\cite{BESIII:2020nme,lu2020online,zhang2022suppression}. The cylindrical core of the BESIII detector covers 93\% of the full solid angle and consists of a helium-based multilayer drift chamber~(MDC), a time-of-flight
system~(TOF), and a CsI(Tl) electromagnetic calorimeter (EMC), which are all enclosed in a superconducting solenoidal magnet providing a 1.0~T magnetic field. The solenoid is supported by an octagonal flux-return yoke with resistive plate counter muon identification modules interleaved with steel. The charged-particle momentum resolution at $1~{\rm GeV}/c$ is
$0.5\%$, and the ${\rm d}E/{\rm d}x$ resolution is $6\%$ for electrons from Bhabha scattering. The EMC measures photon energies with a resolution of $2.5\%$ ($5\%$) at $1$~GeV in the barrel (end cap) region. The time resolution in the plastic scintillator TOF barrel region is 68~ps, while that in the end cap region was 110~ps. The end cap TOF system was upgraded in 2015 using multigap resistive plate chamber technology, providing a time resolution of 60~ps, which benefits 85\% of the data used in this analysis~\cite{Li:2017jpg,Guo:2017sjt,Cao:2020ibk}.

Monte Carlo (MC) simulated data samples produced with a {\sc geant4}-based~\cite{GEANT4:2002zbu} software package, which includes the geometric description of the BESIII detector and the detector response, were used in this study to determine detection efficiencies and estimate backgrounds. A signal MC sample of the decay $D^+\to\gamma e^+\nu_e$ was simulated according to Eq.~(\ref{eq:radlepdec}), where the form factors were taken from Ref.~\cite{Yang:2016wtm}. The minimum energy of the radiative photon was set at 10~MeV to avoid the infrared divergence for soft photons.
In the simulation, we modeled the beam energy spread and initial state radiation (ISR) in the $e^+e^-$ annihilations with the generator {\sc kkmc}~\cite{Jadach:1999vf,Jadach:2000ir}. The inclusive MC sample encompassed the production of $D\bar{D}$ pairs (including quantum coherence for the neutral $D$ channels), non-$D\bar{D}$ decays of the $\psi(3770)$, ISR production of the $J/\psi$ and $\psi(3686)$ states, and continuum processes incorporated in {\sc kkmc}~\cite{Jadach:1999vf,Jadach:2000ir}. All particle decays were modeled with {\sc evtgen}~\cite{Lange:2001uf,Ping:2008zz} using BFs 
either taken from the Particle Data Group~\cite{pdg}, when available, or otherwise estimated with {\sc lundcharm}~\cite{Chen:2000tv,Yang:2014vra}. Final state radiation from charged final state particles was incorporated using the {\sc photos} package~\cite{Barberio:1990ms}. 

\section{Initial data analysis}~\label{sec:inisel}
Taking advantage of the $D^+D^-$ pair production in $e^+e^-$ annihilation at $\sqrt{s}=3.773$ GeV, we employed a double-tag (DT) analysis method~\cite{MARK-III:1985hbd} to constrain the kinematics of $D^+\to\gamma e^+\nu_e$ and suppress non-$D^+D^-$ backgrounds. The $D^-$ meson was firstly reconstructed through six specific hadronic decay modes, $K^+\pi^-\pi^-$, $K^+\pi^-\pi^-\pi^0$, $K^0_S\pi^-$, $K^0_S\pi^-\pi^0$, $K^0_S\pi^+\pi^-\pi^-$ and $K^+K^-\pi^-$. Here, $K^0_S$ was reconstructed with a $\pi^+\pi^-$ pair, and $\pi^0$ was reconstructed from two photons. These decay modes are referred to as tag modes, and the selected events are called  single-tag (ST) events. In the presence of ST $D^-$ meson, the signal decay $D^+\to \gamma e^+\nu_e$ was searched for within its recoil system, with selected events referred to as DT events. The BF of the signal decay was determined from
\begin{linenomath*}
\begin{equation}
\mathcal{B}_{\rm sig}=\frac{N_{\rm DT}}{\sum_i\left(N_{\rm ST}^{i}\epsilon_{\rm DT}^i/\epsilon^i_{\rm ST}\right)},
\label{eq:stbf}
\end{equation}
\end{linenomath*}
where $N_{\rm DT}$ represents the sum of DT yields for all tag modes, and $N^i_{\rm ST}$, $\epsilon^i_{\rm ST}$, and $\epsilon^i_{\rm DT}$ are the ST yield, ST efficiency, and DT efficiency for the tag mode $i$, respectively.

For reconstructing the ST $D^-$ meson, charged tracks detected in the MDC were required to be within a polar angle ($\theta$) range of $|\rm{cos\theta}|<0.93$, where $\theta$ is defined with respect to the $z$-axis, which is the symmetry axis of the MDC. For charged tracks not originating from $K_S^0$ decays, the distance of closest approach to the interaction point (IP) was required to be less than 10\,cm along the $z$-axis, $V_{z}$, and less than 1\,cm in the transverse plane, $V_{xy}$. Particle identification~(PID) for charged tracks combines measurements of the energy loss in the MDC~(d$E$/d$x$) and the flight time in the TOF to form likelihoods $\mathcal{L}(h)~(h=K,\pi)$ for each hadron $h$ hypothesis. The charged kaons or pions not originating from $K_S^0$ were required to satisfy $\mathcal{L}(K)>\mathcal{L}(\pi)$ or $\mathcal{L}(\pi)>\mathcal{L}(K)$, respectively.

The $K_{S}^0$ candidates were reconstructed from two oppositely charged tracks satisfying $V_{z}<$ 20~cm. The two charged tracks were assigned as $\pi^+\pi^-$ without imposing further PID criteria. They were constrained to originate from a common vertex and required to have an invariant mass in the range of (487, 511)~$\mev/c^2$ corresponding to approximately $\pm3\sigma$ of the detection resolution. The decay length of the $K^0_S$ candidate was required to be greater than twice the vertex resolution away from the IP. 

Photon candidates were identified using showers in the EMC. The deposited energy of each shower was required to exceed 25~MeV in the barrel region ($|\cos \theta|< 0.80$) and 50~MeV in the end cap region ($0.86 <|\cos \theta|< 0.92$). To exclude showers that originated from charged tracks, the angle subtended by the EMC shower and the position of the closest charged track at the EMC was required to exceed 10 degrees as measured from the IP. To suppress electronic noise and showers unrelated to the event, the difference between the EMC time and the event start time was required to be within [0, 700]\,ns. The $\pi^0$ mesons were reconstructed using a kinematic fit from all combinations of two photon candidates by constraining their invariant mass to the nominal $\pi^0$ mass~\cite{pdg}. The invariant mass of two photons before the kinematic fit was required to be in the range of (115, 150)~$\mev/c^2$. The combinations with both photons from end cap EMC regions are rejected because of poor resolution.

After selecting the component particles, we reconstruct the ST $D^-$ meson using their combinations. Two kinematic variables, the energy difference $\Delta E$, and the beam-constrained mass $M_{\rm BC}$ were calculated for signal extraction as follows:
\begin{linenomath}
\begin{align}
\Delta E&=E_{D^-}-E_{\rm beam},\\
M_{\rm BC}&=\sqrt{E^2_{\rm beam}/c^4-\left|\vec{p}_{D^-}\right|^2/c^2},
\label{eq:mbc}
\end{align}
\end{linenomath}
where $E_{\rm beam}$ is the beam energy, and $\vec{p}_{D^-}$ and $E_{D^-}$ are the momentum vector and energy of $D^-$ candidate in the $e^+e^-$ rest frame, respectively. If multiple combinations were present in an event, the combination with the smallest $|\Delta E|$ was retained for further analysis for each tag mode and charge. To suppress combinatorial backgrounds, mode-dependent requirements on $\Delta E$ were imposed on the accepted candidates within an approximate range of $\pm3.5\sigma$ for the detection resolution, as listed in Table~\ref{tab:ST}. 

\vspace{1em}
\begin{center}
\footnotesize
\tabcaption{$\Delta E$ requirements, ST yields in data, ST and DT efficiencies for each tag mode in data. Note the efficiencies do not include the BFs of subsequent $\pi^0$ and $K^0_S$ decays.}
\label{tab:ST}
\setlength\tabcolsep{0pt}
\begin{tabular*}{\linewidth}{@{\extracolsep{\fill}} l|cccc }
\hline\hline
Tag mode & $\Delta E$ (MeV) & $N_{\rm{ST}}^{i}\ (\times10^3)$ & $\epsilon_{\rm{ST}}^{i}\ (\%)$ & $\epsilon_{\rm{DT}}^{i}\ (\%)$ \\
\hline
$K^+\pi^-\pi^-$ & $(-25,\ 24)$ & $5552.8\pm2.5$ & $51.10$ & 10.78 \\
$K^0_S\pi^-$ & $(-25,\ 26)$ & $656.5\pm0.8$ & $51.42$ & 10.69\\
$K^+\pi^-\pi^-\pi^0$ & $(-57,\ 46)$ & $1723.7\pm1.8$ & $24.40$ & 5.42 \\
$K^0_S\pi^-\pi^0$ & $(-62,\ 49)$ & $1442.4\pm1.5$ & $26.45$ & 5.57\\
$K^0_S\pi^-\pi^-\pi^+$ & $(-28,\ 27)$ & $790.2\pm1.1$ & $29.59$ & 6.62\\
$K^+K^-\pi^-$ & $(-24,\ 23)$ & $481.4\pm0.9$ & $40.91$ & 8.96\\
\hline\hline
\end{tabular*}
\end{center}
\vspace{1em}

The ST yields were obtained through maximum likelihood fits to the $M_{\rm BC}$ distributions after applying the aforementioned requirements, as shown in Figure~\ref{fig:stfit}. In each fit, the signal was described by the shape obtained from MC simulation, convolved with a double-Gaussian function to take into account resolution differences between data and MC simulation results. The background was described by an ARGUS function~\cite{ARGUS:1990hfq} with the endpoint fixed at the beam energy and other parameters left free. A signal region $M_{\rm BC}\in(1.863,\ 1.877)\ \gev/c^2$, corresponding to approximately $\pm3\sigma$ of the detection resolution, was set to further reject the backgrounds. The ST efficiencies were estimated by analyzing the inclusive MC sample with the same method as in data, dividing the fitted yields by the numbers of generated events. Table~\ref{tab:ST} summarizes the ST yields in data and ST efficiencies for each tag mode.

\begin{center}
\includegraphics[width=0.49\textwidth]{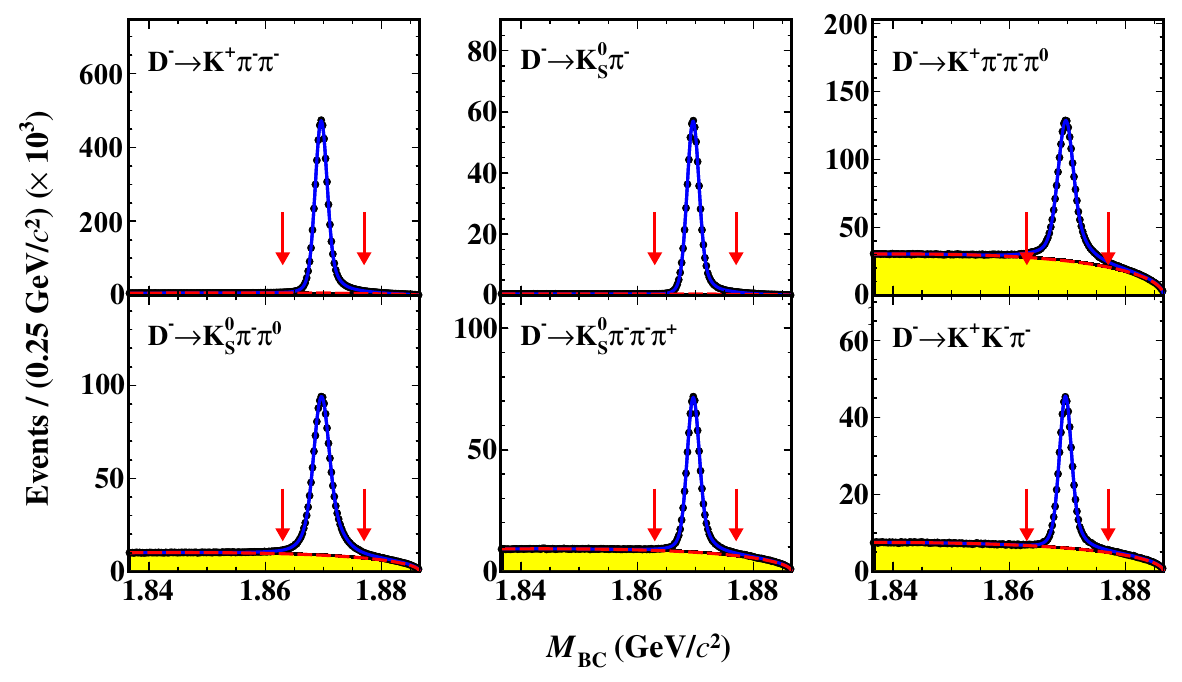}
\figcaption{Fits to the $M_{\rm BC}$ distributions of ST $D^-$ candidates for each tag mode in data. The black dots with error bars represent data. The blue curves are the fit results. The red dotted curves are the fitted combinatorial background shapes. The yellow histograms are the MC simulated background scaled to the luminosity of data. The red arrows indicate signal region.}
\label{fig:stfit}
\end{center}

The signal candidates for $D^+\to\gamma e^+\nu_e$ were selected from the remaining charged tracks and showers, excluding those associated with the ST $D^-$ meson. We required only one track left with charge opposite to that of the ST $D^-$ meson, following the same selection criteria as those in the tag side. This track was required to be identified as a positron combining the PID information from $\mathrm{d}E/\mathrm{d}x$, the TOF system, and the EMC, satisfying $\mathcal{L}(e)>0.001$ and $\mathcal{L}(e)/[\mathcal{L}(e)+\mathcal{L}(\pi)+\mathcal{L}(K)]>0.8$ for the combined confidence levels. We also required that at least one photon candidate satisfied the same selection criteria used in the tag side. If multiple photon candidates passed the selection, the most energetic one was chosen as the signal radiative photon. To improve the kinematic resolution of the positron, final-state-radiation recovery was performed by adding the momenta of any remaining photon candidates back to the reconstructed momentum of positron candidate if their opening angles with respect to the IP were less than $5^\circ$.

To derive information about the missing neutrino, a kinematic observable $U_{\rm miss}$ was defined as
\begin{equation}
    U_{\rm miss}=E_{\rm miss}-\left|\vec{p}_{\rm miss}\right|c,
\end{equation}
where $E_{\rm miss}$ and $\vec{p}_{\rm miss}$ represent the total energy and momentum vector of all missing particles in the event. The missing energy was calculated as $E_{\rm miss}=E_{\rm beam}-E_{e^+}-E_{\gamma}$, where $E_{e^+}$ and $E_{\gamma}$ are the energies of signal positron and photon, respectively. The missing momentum was calculated as $\vec{p}_{\rm miss}=\vec{p}_{D^+}-\vec{p}_{e^+}-\vec{p}_{\gamma}$, where $\vec{p}_{D^+}$, $\vec{p}_{e^+}$ and $\vec{p}_{\gamma}$ are the momenta of signal $D^+$ meson, positron, and photon, respectively. The $D^+$ momentum was further constrained as $\vec{p}_{D^+}=-\frac{\vec{p}_{\rm ST}}{\left|\vec{p}_{\rm ST}\right|}\sqrt{E^2_{\rm beam}-m^2_{D^+}}$, where $\vec{p}_{\rm ST}$ is the momentum of ST $D^-$ meson. The consequent analysis was performed within $U_{\rm miss}\in(-0.2,\ 0.2)$ GeV.

We also explored $D^+\to\pi^0e^+\nu_e$ as a validation channel in this study. For reconstructing $D^+\to\pi^0e^+\nu_e$ events, additional DT selection criteria were applied after those mentioned above. At least one $\pi^0$ candidate was required, and the one with minimum $\chi^2$ from the kinematic fit was chosen as the signal $\pi^0$ candidate. Another kinematic observable, $U'_{\rm miss}$, was defined as
\begin{equation}
    U'_{\rm miss}=E'_{\rm miss}-\left|\vec{p'}_{\rm miss}\right|c,
\end{equation}
where $E'_{\rm miss}=E_{\rm beam}-E_{e^+}-E_{\pi^0}$ and $\vec{p'}_{\rm miss}=\vec{p}_{D^+}-\vec{p}_{e^+}-\vec{p}_{\pi^0}$ were calculated using the energy and momentum of the signal $\pi^0$ candidate, respectively. 

Figure~\ref{fig:before}(a) presents the $U_{\rm miss}$ distributions for both data and inclusive MC samples after applying the above selections. The major backgrounds are from $D^+\to\pi^0e^+\nu_e$, $D^+\to K^0_Le^+\nu_e$, and $D^+\to K^0_S(\to\pi^0\pi^0)e^+\nu_e$. Notably, the $D^+\to\pi^0e^+\nu_e$ background 
\indent\begin{center}
\includegraphics[width=0.4\textwidth]{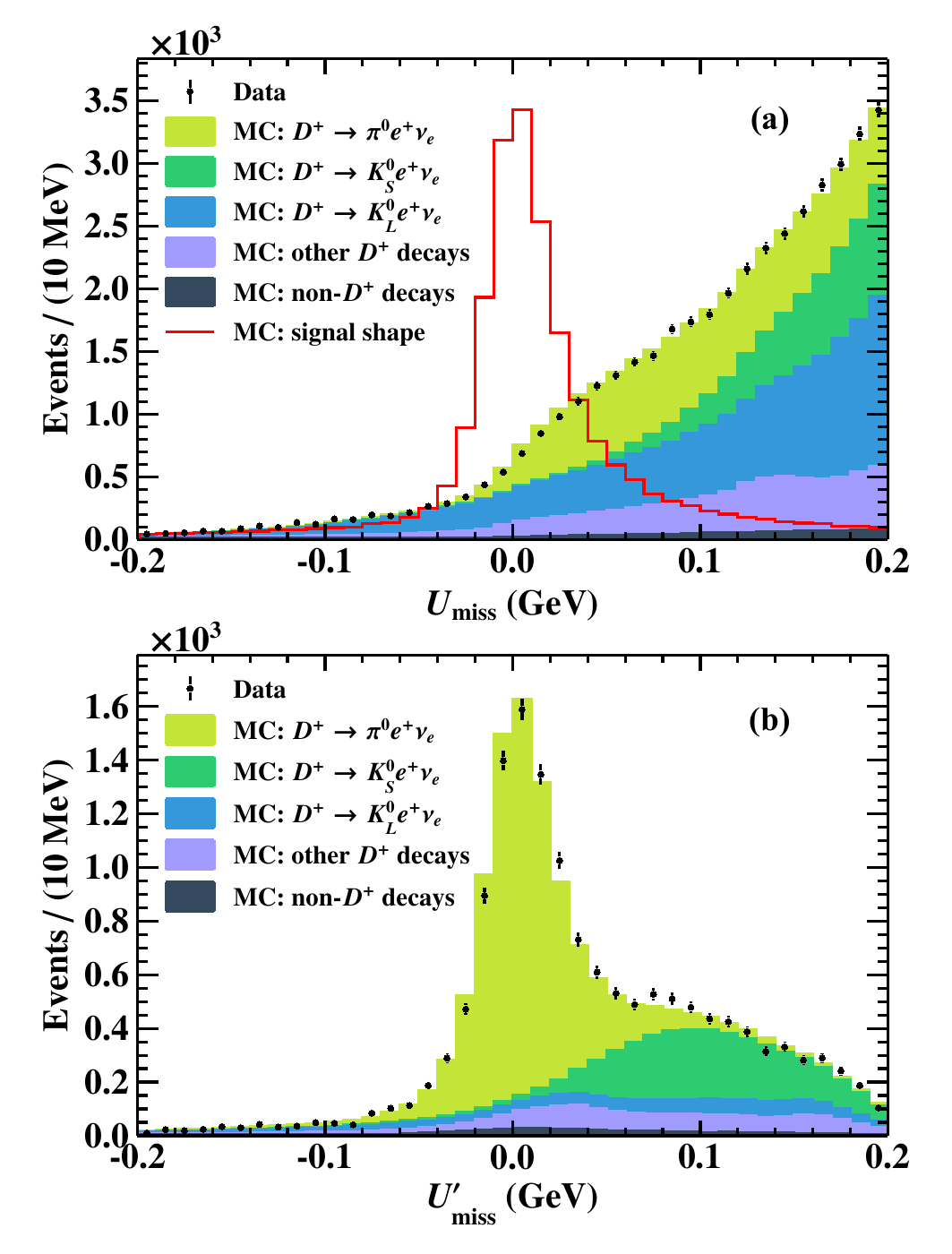}
\figcaption{(a) $U_{\rm miss}$ distribution of the signal decay $D^+\to\gamma e^+\nu_e$ and (b) $U'_{\rm miss}$ distribution of the validation channel $D^+\to\pi^0 e^+\nu_e$ after the DT selection. The black dots with error bars represent data. The histograms with filled colors represent different background components in the inclusive MC sample. The red hollow histogram indicates an arbitrarily scaled signal shape.}
\label{fig:before}
\end{center}
\noindent exhibits a non-trivial shape within the signal region. The high background level makes it difficult to reliably extract the signal yield from the $U_{\rm miss}$ distribution with adequate sensitivity, thereby requiring further signal identification treatments, as explained below. 

\section{Deep-learning-based signal identification}
The main challenge in distinguishing the signals from massive backgrounds lies in extracting distinctive features from the data. Notably, the shower patterns observed in the EMC provide key information for this task. Background events involving a $\pi^0$ typically result in a higher number of photon candidates than the signal, whereas $K^0_L$-induced showers often exhibit more dispersed cluster shapes compared to those from photons~\cite{Li:2024pox}. Additionally, the kinematics of the particles, influenced by the unique phase spaces and dynamics of their decays, offer another aspect of differentiation. Conventional cut-based methods characterize such distinctions via hand-crafted variables derived from practical understanding of the involved physics, such as the number of $\pi^0$-like photon combinations, lateral moment of photon candidates~\cite{BESIII:2017whk}, and $\chi^2$ values from kinematic fits under specific final-state hypotheses~\cite{BESIII:2019pjk}. In contrast, deep learning methods aim to process minimally refined data at the level of fundamental detector responses to the penetrations of the final state particles. A DNN is expected to automatically explore underlying correlations and recognize distinctive information beyond prior knowledge.

In this study, the dataset for DNN training was constructed by randomly shuffling signal and background events sourced from relevant MC samples. The events were categorized into three groups with equal statistics: $D^+\to\gamma e^+\nu_e$ signal, $D^+\to\pi^0 e^+\nu_e$ background, and other inclusive backgrounds. The training dataset contained approximate 3.3 million events in total, with 80\% used for training and the remaining 20\% reserved for validation. The DNN input information from an event included all charged tracks reconstructed in the MDC and isolated showers clustered in the EMC. These particles were organized as a {\it point cloud}~\cite{Qu:2019gqs} structure, i.e., an unordered and variable-sized set of points in a high-dimensional feature space. For each charged track, the features comprised the azimuthal and polar angles in the center-of-mass frame, charge, magnitude of momentum, and parameters characterizing its helical trajectory in the MDC. Additionally, low-level measurements from the MDC, TOF, and EMC were incorporated as implicit information for particle identification and reconstruction quality. For each shower, the features comprised the azimuthal and polar angles, energy deposition, count of fired crystals, time measurement, and parameters describing its expansion scope among nearby crystals. A full list of input features is provided in Table~\ref{tab:feature}; these features were found to be consistent between data and MC simulation results.

\begin{table*}[htbp]
\begin{center}
\footnotesize
\tabcaption{Input particle feature variables and their availability in tracks and showers. If measurements were unavailable from the detector readouts, the default value was set to zero.}
\label{tab:feature}
\setlength\tabcolsep{0pt}
\begin{tabular*}{\linewidth}{@{\extracolsep{\fill}} m{0.2\textwidth}<{\centering}m{0.64\textwidth}>{\centering}m{0.08\textwidth}>{\centering\arraybackslash}m{0.08\textwidth} }
\hline\hline
Variable & Definition & Track & Shower\\
\hline
$\theta$ & azimuthal angle & \checkmark & \checkmark \\
$\phi$ & polar angle & \checkmark & \checkmark \\
$p$ & momentum magnitude & \checkmark & -- \\
\verb|charge| & electric charge & \checkmark & -- \\
$d_{\rho}$ & distance in the $x-y$ plane from the point of closest approach (POCA) of the track to the IP & \checkmark & -- \\
$d_{Z}$ & distance along $z$-axis from the POCA of the track to the IP & \checkmark & -- \\
$\phi_0$ & azimuthal angle of the track at the POCA  & \checkmark & -- \\
$\kappa$ & scaled inverse of the momentum of the track  & \checkmark & -- \\
$\tan\lambda$ & tangent value of the angle of the track at the POCA with respect to the $x-y$ plane  & \checkmark & -- \\
$\chi_{{\rm d}E/{\rm d}x}(e)$ & deviation in ${\rm d}E/{\rm d}x$ measurement of the track from the anticipated positron value  & \checkmark & -- \\
$\chi_{{\rm d}E/{\rm d}x}(\pi)$ & deviation in ${\rm d}E/{\rm d}x$ measurement of the track from the anticipated pion value & \checkmark & -- \\
$\chi_{{\rm d}E/{\rm d}x}(K)$ & deviation in ${\rm d}E/{\rm d}x$ measurement of the track from the anticipated kaon value & \checkmark & -- \\
$\chi_{\rm TOF}(\pi)$ & deviation in TOF measurement of the track from the anticipated pion value  & \checkmark & -- \\
$\chi_{\rm TOF}(K)$ & deviation in TOF measurement of the track from the anticipated kaon value  & \checkmark & -- \\
$Q_{\rm TOF}$ & pulse height of the track in TOF corresponding to its deposited energy  & \checkmark & -- \\
$\beta_{\rm TOF}$ & velocity of the track measured by TOF  & \checkmark & -- \\
$E_{\rm raw}$ & total deposited energy in EMC & \checkmark & \checkmark \\
$E_{\rm seed}$ & deposited energy of the crystal at the shower center & \checkmark & \checkmark \\
$E_{3\times3}$ & sum of energies of $3\times3$ crystal matrices surrounding the shower center & \checkmark & \checkmark \\
$E_{5\times5}$ & sum of energies of $5\times5$ crystal matrices surrounding the shower center & \checkmark & \checkmark \\
$N_{\rm hits}$ & count of fired crystals in the shower & \checkmark & \checkmark \\
\verb|time| & EMC time measurement & \checkmark & \checkmark \\
\verb|secondary| \verb|moment| & shower expansion parameter defined as $\sum_{i}E_{i}r^2_{i}/\sum_{i}E_{i}$ & \checkmark & \checkmark \\
\verb|lateral| \verb|moment| & shower expansion parameter defined as~$\sum_{i=3}^{n} E_{i}r_{i}^2/(E_{1}r_{0}^2+E_{2}r_{0}^2+\sum_{i=3}^{n}E_{i}r_{i}^2)$ & \checkmark & \checkmark \\
$A_{20}$ \verb|moment| & shower expansion parameter defined as $\sum_{i}\frac{E_{i}}{E_{tot}}f_{2,0}(\frac{r_{i}}{R_{0}})$, $f_{2,0}(x)=2x^2-1$ & \checkmark & \checkmark \\
$A_{42}$ \verb|moment| & same but defined as $\left|\sum_{i}\frac{E_{i}}{E_{tot}}f_{4,2}(\frac{r_{i}}{R_{0}})e^{2i\phi}\right|$,~$f_{4,2}(x)=4x^4-3x^2$ & \checkmark & \checkmark \\
\hline\hline
\end{tabular*}
\end{center}
\end{table*}

The architecture of the DNN used in this study largely inherits from the Particle Transformer (ParT)~\cite{qu2022particle}, with adaptations tailored for the BESIII experiment. A key innovation of the ParT is the injection of particle physics knowledge into the Transformer network~\cite{vaswani2017attention}, where a set of pairwise interaction features are calculated using the energy-momentum vectors for each pair of particles and incorporated as biases $\mathbf{U}$ in the self-attention mechanism. This treatment efficiently captures the kinematic relationships between particles, significantly improving the ParT performance. For the BESIII application, we considered its unique properties, such as the purely spherical coordinate system and energy-momentum conservation in 3D space, by designing a new set of pairwise interaction features $\mathbf{U}_{\rm BES}$ defined as
\begin{linenomath*}
\small
\begin{equation}
\mathbf{U}_{\rm BES}=\left\{
\begin{aligned}
&1-\cos\theta_{a,b},\\
&{\rm min}\left(E_a,\,E_b\right)\left(1-\cos\theta_{a,b}\right),\\
&{\rm min}\left(E_a,\,E_b\right)/\left(E_a+E_b\right),\\
&\left(E_a+E_b\right)^2-\lVert\vec{p}_a+\vec{p}_b\rVert^2,
\end{aligned}
\right.
\label{eq:pairwise2}
\end{equation}
\end{linenomath*}
where $E_a$ and $E_b$ are the energies of the two particles, $\vec{p}_a$ and $\vec{p}_b$ are their momentum vectors, and $\theta_{a,b}$ is the opening angle between these vectors in the center-of-mass frame. For the model hyperparameters, this adaptation includes four particle attention blocks and two class attention blocks. It uses a particle embedding encoded from the input particle features via a 3-layer multilayer perceptron with $128,\ 256,\ 128$ nodes in each layer and an interaction embedding encoded from the input pairwise features using a 4-layer pointwise 1D convolution with $64,\ 64,\ 64,\ 8$ channels. A dropout rate of 0.1 was applied to prevent overfitting. 

The training of the DNN was performed on the aforementioned dataset for 50 epochs with a batch size of 512 and an initial learning rate of 0.001. An additional treatment was employed during training to address the potential correlation between the DNN output and the $U_{\rm miss}$ distribution, which could inadvertently create signal-like structures in the $U_{\rm miss}$ distribution of remaining background events after applying the selection criteria on the DNN output~\cite{Dolen:2016kst,Englert:2018cfo,Kasieczka:2020yyl,Kitouni:2020xgb}. To decouple the DNN output from the $U_{\rm miss}$ observable, we employed an iterative approach by assigning a weight $\omega(U_{\rm miss})$ to each background event which adjusts its contribution to the loss function. Events with higher weights have a greater impact on the loss function, which means that they are better classified during training. The weight $\omega_i^j(U_{\rm miss}^j)$ for an event $j$ in the $i$-th iteration is calculated as
\begin{linenomath*}
\begin{equation}
\omega_i^j(U_{\rm miss}^j)=\omega_{i-1}^j(U_{\rm miss}^j)\left[ \left.\frac{p^{\rm BKG}_{i-1}(U_{\rm miss})}{p^{\rm BKG}_{\rm orig}(U_{\rm miss})}\right|_{U_{\rm miss}=U_{\rm miss}^j} \right],
\label{eq:planing}
\end{equation}
\end{linenomath*}
where $\omega_0^j(U_{\rm miss}^j)=1$, $p^{\rm BKG}_{i-1}(U_{\rm miss})$ represents the normalized probability density function for the remaining background shape after the $(i-1)$-th iteration, and $p^{\rm BKG}_{\rm orig}(U_{\rm miss})$ denotes the original background shape before applying the DNN output veto. For each iteration, the DNN was re-trained, and both $p^{\rm BKG}_{i-1}(U_{\rm miss})$ and $\omega_i^j(U_{\rm miss}^j)$ were updated. Intuitively, the process converged when $p^{\rm BKG}_{i-1}(U_{\rm miss})$ approached $p^{\rm BKG}_{\rm orig}(U_{\rm miss})$, resulting in a trivial background shape that facilitates signal extraction. In practice, as shown in Figure~\ref{fig:iter}, this iterative approach partially corrects the $D^+\to\pi^0 e^+\nu_e$ background shape while successfully retrieves the original one for other inclusive background shape. This result demonstrates both its effectiveness and limitations in addressing the mass correlation issue. Furthermore, another machine-learning technique called {\it model ensemble} was employed. With some randomness factors incorporated in the training, such as network initialization, batch processing sequence, and dropout~\cite{Srivastava:2014kpo}, we trained a total of 20 DNNs in parallel in each iteration. The outputs of these DNNs were averaged for each event during inference, creating an ensemble DNN that offers better robustness and generalization than any single DNN model.

\begin{center}
\includegraphics[width=0.49\textwidth]{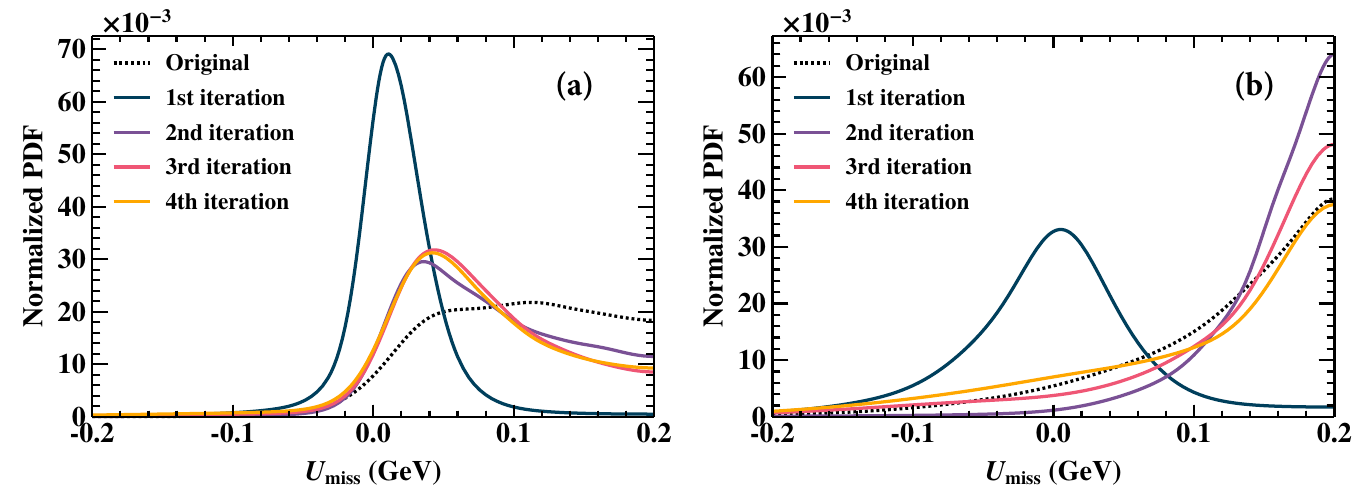}
\figcaption{Normalized (a) $D^+\to\pi^0 e^+\nu_e$ and (b) other inclusive background shapes in the $U_{\rm miss}$ distribution during the iteration process.}
\label{fig:iter} 
\end{center}

The DNN output assigns three scores in the range $[0,\,1]$ to each event, reflecting the probabilities of the event belonging to the signal, $D^+\to\pi^0 e^+\nu_e$ background and other background categories. Given that the sum of the three scores always equals one, only two of them are independent. As the final step in the event selection, we required the score for the $D^+\to\pi^0 e^+\nu_e$ background to be less than 0.15 and that for other backgrounds to be less than 0.05. These cut values were optimized using the Punzi significance metric, $\epsilon/(1.5 + \sqrt{B})$~\cite{Punzi:2003bu}, where $\epsilon$ represents the signal efficiency and $B$ is the expected background yield. Figure~\ref{fig:after}(a) shows the $U_{\rm miss}$ distribution after implementing the deep learning approach. The backgrounds were reduced by more than two orders of magnitude at the cost of approximately two-thirds of the signal efficiency, significantly enhancing the sensitivity of the signal search. 
\begin{center}
\includegraphics[width=0.37\textwidth]{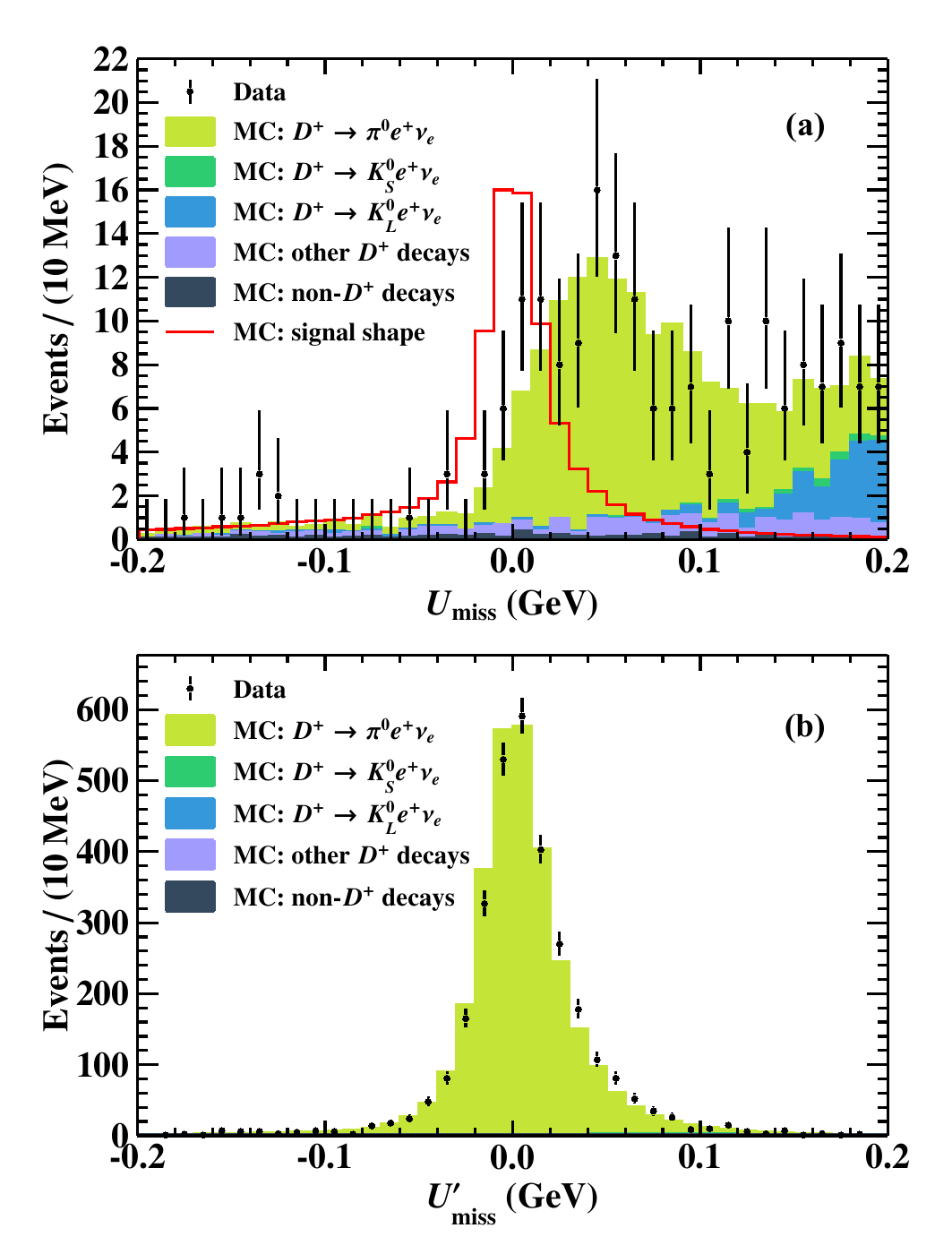}
\figcaption{(a) $U_{\rm miss}$ distribution of the signal decay $D^+\to\gamma e^+\nu_e$ and (b) $U'_{\rm miss}$ distribution of the validation decay $D^+\to\pi^0 e^+\nu_e$ after the DNN vetoes. The black dots with error bars represent data. The histograms with filled colors represent different background components in the inclusive MC sample. The red hollow histogram indicates an arbitrarily scaled signal shape. }\label{fig:after}
\end{center}

\section{Signal extraction}\label{sec:fit}
According to Eq.~(\ref{eq:stbf}), the DT efficiencies for each tag mode, $\epsilon^i_{\rm DT}$, were estimated using signal MC sample; they are summarized in Table~\ref{tab:ST}. To obtain the DT yield $N_{\rm DT}$ in data, a maximum likelihood fit was performed on the $U_{\rm miss}$ distribution combining all the tag modes. The fit included three components, namely the signal, $D^+\to\pi^0 e^+\nu_e$ background, and other backgrounds. The shapes for these components were taken from the corresponding MC simulations, and their yields were all floated in the fit. The fit resulted in $N_{\rm DT}=12.2\pm8.6$ and $\mathcal{B}(D^+\to\gamma e^+\nu_e)=(0.54\pm0.38)\times10^{-5}$, as shown in Figure~\ref{fig:fit}(a), where the uncertainties are statistical only. Given that no significant signal was observed, an upper limit on the signal BF was set after accounting for all systematic uncertainties.

\begin{center}
\includegraphics[width=0.4\textwidth]{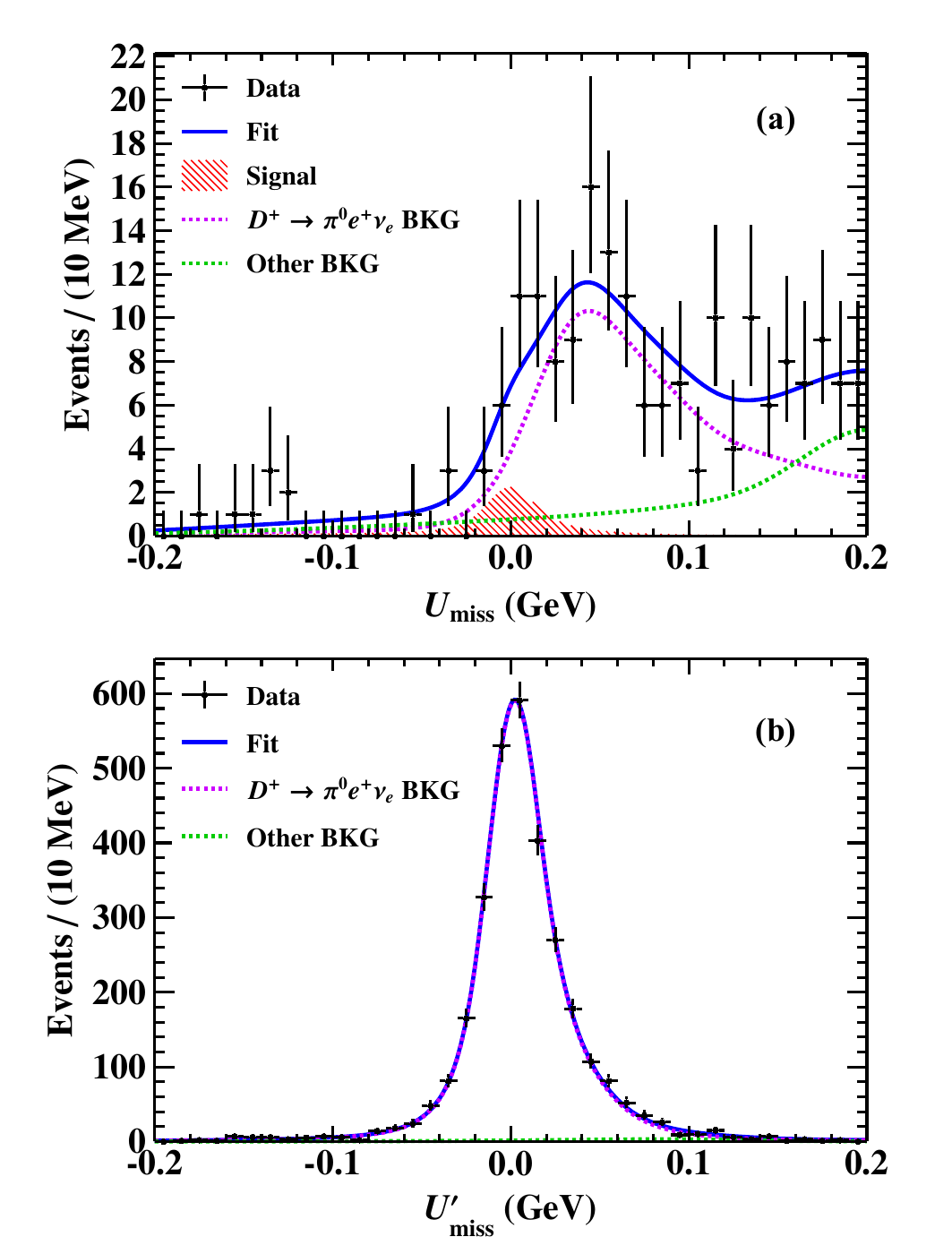}
\figcaption{Fits to (a) the $U_{\rm miss}$ distribution of the signal decay $D^+\to\gamma e^+\nu_e$ and (b) the $U'_{\rm miss}$ distribution of the validation decay $D^+\to\pi^0 e^+\nu_e$. The points with error bars represent data. The blue line represents the fitted curve. The red dashed area represents signal component and other dashed lines represent background components.}\label{fig:fit}
\end{center}

\section{Validation}\label{sec:val}
A common concern in adopting the deep learning method in particle physics experiments is the coupling between potential analysis biases and the sophisticated DNN structure, which makes these biases difficult to validate, quantify, and calibrate. The extensive and diverse datasets collected at BESIII facilitate various control samples in data to investigate these issues. In this study, our primary background process $D^+\to\pi^0e^+\nu_e$ serves as a natural control sample. By partially reversing the DNN vetoes, i.e., to require the score for the $D^+\to\pi^0 e^+\nu_e$ background to be \textit{greater} than 0.15 while keeping the score for other backgrounds below 0.05, we can isolate a $D^+\to\pi^0e^+\nu_e$ sample with a purity exceeding 98\% and measure its absolute BF. Figures~\ref{fig:before}(b) and~\ref{fig:after}(b) show the $U'_{\rm miss}$ distributions before and after applying the partially reversed DNN vetoes, with significant subtraction of backgrounds other than $D^+\to\pi^0e^+\nu_e$. 

To measure the absolute BF of $D^+\to\pi^0e^+\nu_e$, we used the value of $\epsilon^i_{\rm DT}$ estimated from MC simulations and extract $N_{\rm DT}$ via the maximum likelihood fit shown in Figure~\ref{fig:fit}(b). In the fit, the $D^+\to\pi^0e^+\nu_e$ signal was modeled using the MC simulated shape convolved with a Gaussian resolution function. The other background component was described using the inclusive MC shape with the yield floated. The final fit to all data resulted in $N_{\rm DT}=3004\pm59$ and $\mathcal{B}(D^+\to\pi^0 e^+\nu_e)=(3.629~\pm0.071)~\times10^{-3}$, where the uncertainty was statistical only. This result is consistent with previous BESIII measurement $(3.63~\pm0.08_{\rm stat.}~\pm0.05_{\rm syst.})\times10^{-3}$~\cite{BESIII:2017ylw}, based on about one sevenths of full data, thereby validating the reliability of our DNN method. The residual deviations were considered a source of systematic uncertainty, as detailed in the next section.

\section{Systematic uncertainties}
Systematic uncertainties in BF determination stem from various sources; many of them are related to the ST selection being canceled, benefiting from the DT technique. The remaining systematic sources can be grouped into two categories: multiplicative uncertainties, which affect the efficiency, and additive uncertainties, which impact the signal yield. These two types of uncertainties contribute to the upper limit of the BF through different approaches.

The multiplicative uncertainty sources include the ST yield, positron tracking, positron PID, photon reconstruction, MC model, and DNN vetoes, as summarized in Table~\ref{tab:sys}. The uncertainty due to the ST yield was estimated by varying the descriptions of signal and background components in the $M_{\rm BC}$ fits and was set to 0.3\%. A control sample of radiative Bhabha scattering events was used to study the position tracking and PID uncertainties, and the values for both were set to 1.0\%~\cite{BESIII:2019qci}. The photon reconstruction uncertainty was set to 1.0\% using a control sample of $J/\psi\to\pi^+\pi^-\pi^0,\, \pi^0\to\gamma\gamma$~\cite{BESIII:2011ysp}. The uncertainty related to MC modeling was defined as the difference between the signal efficiencies of the nominal MC model and an alternative single-pole model~\cite{Yang:2014rna}; it was set to 9.1\%.

Special consideration was given to systematic uncertainties related to the DNN vetoes~\cite{Shanahan:2022ifi}. Following some frontier surveys~\cite{pdg,Chen:2022pzc}, we consider two primary uncertainty sources: {\it model uncertainty} and {\it domain shift}. Model uncertainty arises from our limited knowledge of the best model and was quantified by the relative shift in signal efficiency across the 20 DNNs trained during the model ensemble process. The standard deviation of these efficiencies, divided by their average, 5.3\%, was taken as the corresponding systematic uncertainty. Domain shift describes the mismatch between datasets for training and inference, reflecting potential data-MC inconsistencies. This uncertainty was evaluated using the control sample of $D^+\to\pi^0 e^+\nu_e$ described in Section~\ref{sec:val}, with the deviation between our BF measurement and the previous BESIII measurement~\cite{BESIII:2017ylw} calculated to be 2.2\%, combining the center value shift and $1\sigma$ statistical uncertainty. In total, the multiplicative systematic uncertainty was determined to be 10.9\% by adding the above sources in quadrature.

\vspace{1em}
\begin{center}
\footnotesize
\tabcaption{The multiplicative systematic uncertainties.}
\label{tab:sys}
\setlength\tabcolsep{0pt}
\begin{tabular*}{\linewidth}{@{\extracolsep{\fill}} lc }
\hline \hline
Source                   & Size (\%) \\
\hline
ST yield                  & 0.3 \\
Positron tracking         & 1.0 \\
Positron PID              & 1.0 \\
Photon reconstruction     & 1.0 \\
MC model                  & 9.1 \\
DNN vetoes      & 5.8 \\
\hline
Total                     & 10.9 \\
\hline \hline
\end{tabular*}
\end{center}
\vspace{1em}

The additive systematic uncertainties arise from the signal, $D^+\to\pi^0 e^+\nu_e$ background, and other background shapes used in the DT fit. These effects were evaluated by testing alternative descriptions of the shapes. One variation is in using parameterized functions, where the signal and $D^+\to\pi^0 e^+\nu_e$ background shapes were modeled by double-Gaussian functions, and the other background shape was modeled a third-order Chebyshev polynomial. Another change was the use of MC-simulated shapes before the final iteration of the mass decorrelation process as well as across the 20 DNN predictions in the model ensemble process.

\section{Results}
Given that no significant signal was observed in data, the upper limit on $\mathcal{B}(D^+\to\gamma e^+\nu_e)$ was set using a Bayesian method described in Ref.~\cite{Stenson:2006gwf}. We performed a series of maximum likelihood fits as described in Section~\ref{sec:fit}, with the signal yield $N_{\rm DT}$ fixed to each scanned value. The corresponding maximum likelihood values were used to construct a discrete likelihood distribution, $\mathcal{L}(\mathcal{B}_{\rm sig})$. The systematic uncertainties were incorporated in two steps: first, the additive uncertainties were addressed by varying the DT fit method, with the most conservative upper limit retained. Next, the multiplicative uncertainties were considered by smearing the likelihood distribution as follows~\cite{Liu:2015uha}
\begin{equation}
\mathcal{L}'(\mathcal{B}_{\rm sig})\propto\int^{1}_{0}\mathcal{L}\Big(\mathcal{B}_{\rm sig}\cdot\frac{\epsilon}{\epsilon_{0}}\Big)e^{-\frac{(\epsilon-\epsilon_{0})^{2}}{2\sigma_{\epsilon}^{2}}}\mathrm{d}\epsilon,
\label{eqn:ulsmear}
\end{equation}
where $\epsilon_{0}$ is the nominal signal efficiency, and $\sigma_{\epsilon}$ is the multiplicative uncertainty corresponding to the efficiency value. By integrating the $\mathcal{L}'(\mathcal{B}_{\rm sig})$ curve up to 90\% of the area for $N_{\rm DT}>0$ and calculating the corresponding BF using Eq.~(\ref{eq:stbf}), we set the upper limit on the BF of $D^+\to\gamma e^+\nu_e$ at 90\% C.L. to be $1.2\times10^{-5}$, as shown in Figure~\ref{fig:ul},

\begin{center}
\includegraphics[width=0.4\textwidth]{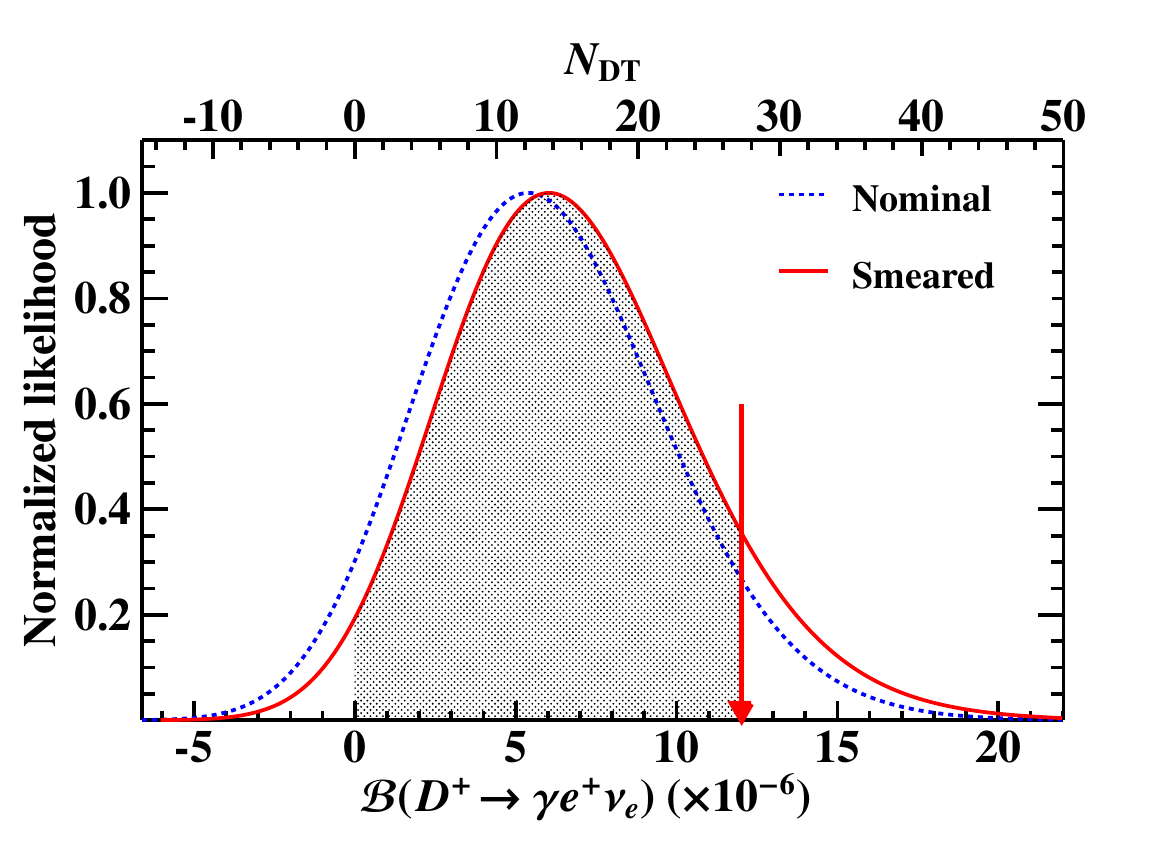}
\figcaption{The normalized likelihood distributions before and after incorporating the multiplicative and additive systematic uncertainties. The blue dashed line represents the initial distribution and the red solid line represents the distribution with systematic uncertainties considered. The red arrow indicates the final upper limit.}
\label{fig:ul}
\end{center}

\section{Summary}
We conducted an improved search for the radiative leptonic decay $D^+\to\gamma e^+ \nu_e$ using 20.3~$\ifb$ of $e^+e^-$ collision data taken at $\sqrt{s}=3.773$ GeV with the BESIII detector. Our investigation does not reveal a significant signal but establishes an upper limit of $1.2\times10^{-5}$ on its BF with a radiative photon energy above 10~MeV, at 90\% C.L.. This result excludes most existing theoretical predictions~\cite{Barik:2009zza,Korchemsky:1999qb,Yang:2014rna,Yang:2016wtm,Lu:2021ttf}, thereby indicating the necessity for further investigation of the decay mechanism. The key advancement of this study is the application of a novel deep learning approach, which effectively distinguishes the signals from massive backgrounds. Our results showcase the potential and broad applicability of deep learning techniques in particle physics research.

\section{Acknowledgment}
{\it The BESIII Collaboration thanks the staff of BEPCII and the IHEP computing center for their strong support.}

%%%%%%%%%%%%%%%%%%%%%%%%%%%%%%%%%%%%%%%%%%%%%%%
\end{multicols}
\vspace{-1mm}
\centerline{\rule{80mm}{0.1pt}}
\vspace{2mm}
\begin{multicols}{2}
\bibliographystyle{apsrev4-2}
\bibliography{bibitem}
\end{multicols}
\end{CJK*}

\end{document}

%% file: authorlist_enzh_2024-11-18.tex
%\author{Author list}
M.~Ablikim(麦迪娜)$^{1}$, M.~N.~Achasov$^{4,c}$, P.~Adlarson$^{76}$, X.~C.~Ai(艾小聪)$^{81}$, R.~Aliberti$^{35}$, A.~Amoroso$^{75A,75C}$, Q.~An(安琪)$^{72,58,a}$, Y.~Bai(白羽)$^{57}$, O.~Bakina$^{36}$, Y.~Ban(班勇)$^{46,h}$
% ...
\\
\vspace{0.2cm}
(BESIII Collaboration)\\
\vspace{0.2cm} {\it
$^{1}$ Institute of High Energy Physics, Beijing 100049, People's Republic of China\\
% ...
$^{4}$ Budker Institute of Nuclear Physics SB RAS (BINP), Novosibirsk 630090, Russia\\
% ...
$^{35}$ Johannes Gutenberg University of Mainz, Johann-Joachim-Becher-Weg 45, D-55099 Mainz, Germany\\
$^{36}$ Joint Institute for Nuclear Research, 141980 Dubna, Moscow region, Russia\\
% ...
$^{46}$ Peking University, Beijing 100871, People's Republic of China\\
% ...
$^{57}$ Southeast University, Nanjing 211100, People's Republic of China\\
$^{58}$ State Key Laboratory of Particle Detection and Electronics, Beijing 100049, Hefei 230026, People's Republic of China\\
% ...
$^{72}$ University of Science and Technology of China, Hefei 230026, People's Republic of China\\
% ...
$^{75}$ University of Turin and INFN, (A)University of Turin, I-10125, Turin, Italy; (B)University of Eastern Piedmont, I-15121, Alessandria, Italy; (C)INFN, I-10125, Turin, Italy\\
$^{76}$ Uppsala University, Box 516, SE-75120 Uppsala, Sweden\\
% ...
$^{81}$ Zhengzhou University, Zhengzhou 450001, People's Republic of China\\
\vspace{0.2cm}
$^{a}$ Deceased\\
% ...
$^{c}$ Also at the Novosibirsk State University, Novosibirsk, 630090, Russia\\
% ...
$^{h}$ Also at State Key Laboratory of Nuclear Physics and Technology, Peking University, Beijing 100871, People's Republic of China\\
% ...
}